\documentclass[pra,reprint,nofootinbib,showpacs,superscriptaddress,twocolumn]{revtex4-1}

\usepackage{graphicx}
\usepackage{hyperref}
\usepackage{amsfonts}
\usepackage{amsmath,amssymb}
\usepackage{bm}
\usepackage{color}

\def\be{\begin{equation}}
\def\ee{\end{equation}}
\def\bea{\begin{eqnarray}}
\def\eea{\end{eqnarray}}
\def\ba{\begin{array}}
\def\ea{\end{array}}
\def\bem{\begin{multline}}
\def\eem{\end{multline}}

\begin{document}

%\title{Convergence of quantum generative adversarial learning}

\title{Quantum generative adversarial learning}

\author{Seth Lloyd}
\affiliation{Massachusetts Institute of Technology, Department of Mechanical Engineering,
77 Massachusetts Avenue, Cambridge, Massachusetts 02139, USA}
\author{Christian Weedbrook}
\affiliation{Xanadu, 372 Richmond Street W, Toronto, Ontario M5V 1X6, Canada}

\date{\today}
%\pacs{03.67.-a, 03.67.Ac}

\begin{abstract}
Generative adversarial networks (GANs) represent a powerful tool for
classical machine learning: a generator tries to create statistics
for data that mimics those of a true data set, while a discriminator
tries to discriminate between the true and fake data.  The learning
process for generator and discriminator can be thought of as an adversarial
game, and under reasonable assumptions, the game converges to
the point where the generator generates the same statistics as
the true data and the discriminator is unable to discriminate
between the true and the generated data.    This paper
introduces the notion of quantum generative adversarial networks (QuGANs),
where the data consists either of quantum states, or of
classical data, and the generator and 
discriminator are equipped with quantum information processors. 
We show that the unique fixed point of the quantum adversarial
game also occurs when the generator produces the same statistics
as the data.  Since quantum systems are intrinsically probabilistic 
the proof of the quantum case is different from -- and 
simpler than -- the classical case.   We show that when
the data consists of samples of measurements made on high-dimensional
spaces, quantum adversarial networks may exhibit an exponential
advantage over classical adversarial networks.   
\end{abstract}

\maketitle

%%%%%%%%%%%%%%%%%%%%

%\magnification=\magstep1
%\baselineskip=16pt
%\hfuzz=6pt
%
%$ $
%
%\rightline{April 3rd, 2018}
%\vskip 1in

%\centerline{\bf Convergence of quantum generative adversarial learning}

%\bigskip

%\centerline{SL for CW}

%\bigskip

\section{Introduction}

%- classical ML/AI/deep learning; applications
%
%- QML
%
%- last sentence of paragraph about GANs: One area of classical machine learning that has yet to find it way into the quantum realm is that of generative adverarial networks (GANs)~\cite{Goodfellow2014}. GANS have risen to prominence due to its XXXX.

In machine learning by generative adversarial networks~\cite{Goodfellow2014}, a generator learns to generate statistics
of data by trying to fool a discriminator into believing that the generated
statistics actually come from the data.   
The discriminator is
presented either with real data, or with data generated
by the generator: her goal is to maximize the probability of
assigning the correct label, real or fake, to data.
The generator is equipped with a random number generator
which he uses to try to produce data that minimizes the
probability of the discriminator assigning the correct label:
his goal is to produce data that matches the statistics of the
true data.  That is, the discriminator and generator are 
adversaries in a machine learning game. 

The endpoint of such an adversarial
game, under reasonable assumptions~\cite{Goodfellow2014}, results in the generator producing data with the true statistics,
and the discriminator having a probability of $1/2$ of discriminating
correctly.   
In practice, adversarial games work well in training the
generator to generate data with the true statistics of the data. This has lead to practical applications such generating photorealistic images~\cite{Salimans2016} and videos~\cite{Mathieu2015}, image super resolution~\cite{Ledig2016}, and image inpainting~\cite{Pathak2016}. This has resulted in significant interest in industries such as driverless cars, finance, medicine and cybersecurity. 
%
%- deep learning~\cite{LeCun2015}
%
In this paper, we introduce quantum generative adversarial networks, or QuGANs, where the discriminator,
the generator, and the system generating the actual data can be
quantum mechanical. Various other such quantum machine learning algorithms exists~\cite{Biamonte2017} and show significant benefits over their classical counterparts. Here we consider three specific QuGAN protocols.   

First, we look at the situation where the system, data, discriminator, and generator,
are all fully quantum mechanical: the data takes the form of an ensemble
of quantum states generated by the system, the generator has access to
a quantum information processor and tries to match that ensemble;
the discriminator can make arbitrary quantum measurements.
In this fully quantum setting, in analog to the classical
result, we will show that the quantum 
discriminator and generator perform convex optimization with
the unique fixed point for the quantum adversarial game. This is the situation where the generator accurately reproduces
the true ensemble of quantum states, and the discriminator
can't tell the difference between the true ensemble and
the generated ensemble.

The quantum adversarial game can be formulated in the language of Nash
equilibria for a process in which the discriminator tries to 
optimize her strategy over a fixed number of trials with the 
generator's strategy fixed. This is followed by the generator trying
to optimize his strategy over a number of trials with the
discriminator's strategy fixed. The endpoint of the game,
with the generator finding the correct statistics and the
discriminator unable to tell the difference between true data
and the generated data, is the unique
Nash equilibrium.   

Second, we look at situations where the real data is generated from
the quantum system by a fixed measurement.
The key point in this setting is that relatively simple
quantum systems can generate data whose statistics -- under reasonable
assumptions of computational complexity -- cannot be generated efficiently by
any classical system equipped with a random number generator.  This
feature of quantum systems is sometimes called quantum supremacy
or quantum advantage~\cite{Preskill2018}. Quantum supremacy implies that
a generator that does not have access to
quantum information processing will in general be unable to match
the statistics of data generated by another quantum system.
Consequently, a classical generator will fail to generate
the correct data, and a classical discriminator -- either classical
or quantum -- can in principle make measurements that will
discriminate between the true data and the generated data.
Whether or not the discriminator can make such measurements
in practice, either systematically or by adaptation,
is an open question. 

Finally, we look at the question of whether a quantum generator
can do better than a classical generator at generating
{\it classical} data.   The ability of quantum information
processors to represent vectors in $N$-dimensional spaces
using $\log N$ qubits, and to perform manipulations of sparse
and low-rank matrices in time $O({\rm poly}(\log N))$ implies
that QuGANs exhibit a potential exponential advantage
over classical GANs when the object of the game is to
reproduce the statistics of measurements made on very
high-dimensional data sets. Note that a more in-depth practical implementation of QuGANs can be found in a companion paper~\cite{Dallaire2018}.

\begin{figure}[th]
\vspace{-0.00cm}
\par
\begin{center}
\includegraphics[width=3.5cm]{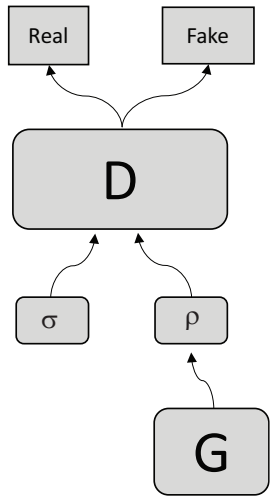}
\end{center}
\par
\vspace{-0.60cm}\caption{Schematic of a general quantum generative adversarial network (QuGAN) protocol. The ultimate goal of this adversarial game is for the discriminator (D) to determine whether the input data is real ($\sigma$) or fake ($\rho$). Here G is the generator which generates fake data hoping to fool the discriminator. We consider a variety of QuGAN situations. Firstly, where the real data is quantum, the generator is quantum (and hence generates fake quantum data), and the discriminator is quantum. Secondly, the real data is quantum, the generator is classical and the discriminator is either classical or quantum. Finally, we consider the case where the real data is purely classical and the generator and discriminator are both quantum.}%
\label{QuGANs}%
\end{figure}

\section{Data quantum, discrimator quantum, generator quantum}

As in the classical adversarial game~\cite{Goodfellow2014}, the quantum adversarial
game is set up as follows.   First the discriminator tries
to improve her strategy, with the generator's strategy fixed.
Then the generator tries to improve his strategy, with the
discriminator's strategy fixed.    The players continue updating
in turn until a fixed point is reached.   We will show
that as long as the generator is producing statistics
that are different from those of the true data,
the discriminator can always adjust her measurement towards
the minimum error discriminating measurement, so that
she succeeds in discriminating true from fake data with
probability $> 1/2$.    Next    
we show that the generator 
can always decrease the probability of success of the discriminator
by moving in a direction that decreases the relative entropy
between the true data and the generated data.    As this is
a convex optimization problem, there is a unique endpoint to
the process, which is where the generator correctly matches the statistics
of the data, and the discriminator is unable to distinguish between
real and fake data with a probability different from $1/2$.

To see that quantum adversarial networks also result in the
generator correctly matching the data, suppose that
the true data is described by an ensemble of states described
by a density matrix $\sigma$,
and the generator generates an ensemble of states with density matrix
$\rho$, cf. Fig.~\ref{QuGANs}.  The discriminator is presented either with a state from the
true ensemble or the generated ensemble, and has to try to
discriminate between them.   First, we assume that the generator
is fixed, and generates $\rho$ for each trial; we then 
train the discriminator to
try to distinguish between $\sigma$ and $\rho$.  Next, we fix
the measurement strategy of the
discriminator and train the generator into trying to adjust
$\rho$ to fool the discriminator.

When $\rho$ is fixed, 
the minimum error measurement to
discriminate between $\sigma$ and $\rho$ is the measurement
with operators $P_+$ and $1-P_+$ that distinguish between the positive and negative part of $\sigma - \rho$~\cite{Helstrom1976}.    
Of course, the discriminator doesn't know the optimal measurement
to begin with. However, she can guess a measurement, and given
feedback for the probabilities of that measurement discriminating
between true and generated data, adjust the measurement by a
process of gradient descent.
The discriminator makes a positive operator valued measurement (POVM)~\cite{Nielsen2000}
$D$ with outcomes $T$ or $F$, $T+F = I$.
The probability that the measurement yields the result $data$
given that the data was indeed selected from the true ensemble
described by $\sigma$ is 
$p(T|data) = {\rm tr} T \sigma $, and the probability that
the measurement yields the result $data$ given that the data was
selected from the generated ensemble is $p(T|G) = {\rm tr} T \rho$.    
$T,F$ are positive operators with $\|T\|_1$, $\|F\|_1\leq 1$.
The set of positive operators with 1-norm less than or equal to 1
is convex.   Accordingly, over many trials, the discriminator
can simply follow the gradient of the function $p(T|data)$
to find the minimum error measurement.

In classical adversarial learning, the discriminator is supplied
with a deep learning network~\cite{LeCun2015} such as a perceptron, whose weights
she adjusts to try to find an optimal measurement.    In the quantum
case we assume that the discriminator is supplied with a quantum
information processor such as a quantum deep learning network
or a quantum circuit that takes as input the quantum state
from the data or the generator and performs the discriminating
measurement. Just as in the classical case, we assume
that the discriminator is able to adjust the weights of her network
to follow the gradient of the $p(T|\sigma)$ for at least some
distance, which may be all the way to the optimal minimum error
measurement.

Once the discriminator has found a good measurement $T,F$ to distinguish
the true from fake data, it is the generator's turn.
The generator tries to adjust the state $\rho$ of the generated
data to maximize $p(T|\rho) = {\rm tr} T\rho$.    The set
of density matrices $\rho$ is convex, and the generator
can follow the gradient of $p(T|\rho)$ to find the state
of the generated data that maximizes the probability of
fooling the discriminator.    Once again, we assume that
the generator possesses a deep quantum network or quantum
circuit whose weights he can adjust to follow the gradient
for at least some distance.  

The adversarial game can be described in the language
of Nash equilibria.   The discriminator's strategy
is given by the measurement operator $T$, and the
generator's strategy is given by the density matrix
$\rho$.    The set of possible positive measurement
operators $T$ made by the discriminator is convex and
compact, as is the set of possible density matrices 
$\rho$ generated by the generator.   Applying the Kakutani
fixed point theorem~\cite{Kakutani1941}, we see that the discrimator-generator
strategy space has at least one fixed point.    In
fact, there is only one fixed point.     Suppose that
the discriminator has found a measurement $T$ such
that $p(T|\sigma) > p(T|\rho)$, so that 
$ {\rm tr} T\sigma - {\rm tr} T\rho > 0$.  If $\rho \neq \sigma$,
such a measurement always exist (e.g., the minimum
error measurement described above).    The
generator can then always increase $p(T|\rho)$
by taking $\rho \rightarrow \alpha(\sigma - \rho)$,
$\alpha > 0$.   Accordingly, the unique Nash equilibrium
occurs when $\rho = \sigma$ and $p(T|\sigma) = p(T|\rho) = 1/2$.
Moreover, as shown above, at each move of the game the
discriminator or generator can move directly towards
this equilibrium by following the gradient of $p(T|\sigma)$
or $p(T|\rho)$ through the convex strategy space.

The result of the
quantum adversarial game is the same as the result of the
classical: the generator learns to generate the data and
the discriminator does no better than chance.   Since
quantum systems are intrinsically probabilistic, however,
the proof of the quantum case is different from -- and indeed,
simpler than -- the classical case.

\section{Data quantum, discriminator quantum or
classical, generator classical}

We now consider the case in which the real data $\sigma$ is being generated
by a quantum system via a fixed measurement, yielding
statistics $p_{true}(x)$ for measurement outcomes $x$.
In this case, quantum supremacy~\cite{Preskill2018} implies that the classical
generator can't efficiently match the statistics of the quantum data.
More precisely the generator is unable
to match his statistics $p_g(x)$ with the true statistics
of the data $p_{true}(x)$, unless he has exponentially scaling resources: $p_g$ is 
bounded away from $p_{true}$ in 1-norm.   Consequently, there
exists a measurement that the discriminator can make that
distinguishes $p_{true}$ from $p_g$ with probability strictly
bounded away from $1/2$.  The minimum error measurement is a projector
onto the positive part of $p_{true} - p_g$, that is, a projector
on the set $X_+$ such that $p_{true}(x) - p_g(x) \geq 0$.    
As long as the discriminator can
find this measurement, then she can win the game.

The key question here is whether the discriminator can
actually find the minimum error measurement or the optimal
probabilistic strategy.   The discriminator's measurement
now corresponds to a POVM with operators $T,F$ that are
diagonal in the measurement basis.    Once again, the
set of such operators is convex,
so under the same assumptions as above on the efficacy
of deep learning in exploring the space of such measurements, 
when the generator's probabilities $p_g$
are fixed, the discriminator can adjust her measurement
strategy to the optimal one, at least in principal.

In the case where the data is generated by a system
that exhibits quantum supremacy, however,
under plausible assumptions of computational complexity,
a classical device can't reproduce the
true probabilities for the data
$p_{true}(x)$. 
In particular, there is no known non-exponential classical
algorithm for determining the optimal measurement to demonstrate
quantum supremacy.  If the discriminator has access to a
quantum information processor to adjust her measurement
strategy, then we conjecture that she
can find the optimal measurement to discriminate between
the quantumly generated data and the classically
generated data.    If the discriminator only has access
to classical information processing, then we conjecture
that she can't determine the optimal measurement.

When the quantum system generating the data does not
exhibit quantum supremacy, as is the case, for example,
for Gaussian continuous-variable systems~\cite{Weedbrook2012}, then both
discriminator and generator can in principle reproduce
the statistics of the data using classical methods.
They can search through possible Gaussian states and
measurements using classical computation, and the
adversarial quantum learning game will in general
lead to an equilibrium where the generator successfully
generates the statistics of the Gaussian quantum data.

\section{Data classical, discriminator and generator quantum}

Now suppose that the data is purely classical, for example,
a set of images taken from the internet, or sequences of prices
of stocks on the stock market, but the generator
and discriminator have access to quantum information processing.
Are there classical data sets for which the quantum adversarial
game is more efficient than the classical one?   Here, `more
efficient' means either that the quantum game converges faster,
or that it uses many fewer resources.
When the data set is classical, no guarantee of quantum supremacy applies.

The ability of quantum information processors to represent
$N$-dimensional vectors using $\log N$ qubits, and to perform linear algebra
on those vectors in time $O({\rm poly}(\log N))$, implies that 
quantum information processors might indeed be able to 
provide a highly compressed version of generative
adversarial learning tasks.    Suppose that the underlying
data consists of $M$ normalized vectors $\vec v_j$ in an $N$-dimensional
real or complex space, so that the (normalized) covariance matrix of the data is
$C= (1/M) \sum_j \vec v_j \vec v_j^\dagger$.    A quantum information
processor can represent those vectors by quantum states
$|v_j\rangle$ over $\log N$ qubits, and the normalized covariance matrix
of the data is equal to the density matrix $\rho = (1/M)
\sum_j |v_j\rangle\langle v_j|$.     Suppose that the
actual observed data consists of the expectation values and
higher moments of a relatively small number $r$
of sparse or low-rank Hermitian matrices $R_\ell$. The goal of
both the classical and quantum generative adversarial
games is to reproduce the statistics of the observed data.

Clearly, the classical generator can produce the observed
data by performing gradient descent in the convex set of normalized
covariance matrices, a task that takes time $O(N^2)$.
If $N$ is large, e.g., $N= 10^{12}$, then the time to
perform this convex optimization is prohibitively large,
$O(10^{24})$ steps.    By contrast, a quantum generator
can represent candidate covariance matrices using
$O(\log N)$ qubits, and evaluate the statistics generated
by such a candidate covariance matrix using $O({\rm poly}(\log N))$
quantum logic operations.   The gradients the quantum
device must follow to try to reproduce the moments 
${\rm tr} R^k_\ell \rho$ are simply
$\partial {\rm tr} R^k_\ell \rho /\partial \rho = R^k_\ell,$
which are themselves sparse or low-rank Hermitian matrices.

The quantum system can follow these gradients efficiently.
Assume for the moment that the $R^k_\ell$ are positive.
If a positive matrix $R$ is low-rank, then methods of density matrix
exponentiation~\cite{Lloyd2014} allow one to 
implement $\rho \rightarrow \rho + \alpha R$ by the
modified swap operations of~\cite{Lloyd2014}.
If they are sparse, then we can use the methods of~\cite{Harrow2009}
to implement $R^{1/2}$ and to construct the density matrix
$R/{\rm tr} R$.   An infinitesimal swap then yields
$\rho \rightarrow \rho + \alpha R$.   That is,      
a quantum generator with $O(\log N)$ qubits can follow
the gradients of the moments of the observed operators
in time $O({\rm poly}(\log N))$.

By contrast, a classical generator that tries to follow
the gradients of the observables explicitly by gradient
descent on the set of covariance operators takes
$O(N^2)$ bits and time $O(N^2)$. Of course, it
may be that a much smaller deep classical network such
as a perceptron or Boltzmann machine may be
able to perform such optimization {\it implicitly}.
Whether a deep network of size $O({\rm poly}(\log N))$
can in fact reproduce the statistics of operators
sampled from a very high dimensional space is an 
open question, which could be tested by direct
numerical experiment. As in the classical case, the analysis of convergence for QuGANs assumes that quantum networks do
indeed have enough flexibility to track the necessary gradients.
In a companion paper~\cite{Dallaire2018}, the authors analyze the ability of such networks
to track gradients and show in the case of small networks that
they can do so effectively. The corresponding result for
large quantum generative networks will have to be verified
directly on quantum information processors.

\section{Discussion and Conclusion}\label{sec:con}

Future work will entail running QuGAN simulations on quantum software packages such as Strawberry Fields~\cite{Killoran2018}. For example, in the classically simulable case of Gaussian states~\cite{Weedbrook2012}, we can simulate both the fully quantum case, where the data, discriminator,
and generator have access to arbitrary Gaussian processes. And also in the quantum-classical case, where the data is generated
by a Gaussian process and the discriminator and generator
are trying to reproduce the statistics of Gaussian measurements.
One could simulate a multi-mode Gaussian process with injected squeezed
states, a unitary mode-mixing transformation, and homodyne
measurements.   

We have shown that in the quantum-quantum case, the unique
Nash equilibrium occurs when the generator reproduces the statistics of the data correctly. In the quantum-classical
case, this is still the unique Nash equilibrium, but quantum supremacy prevents a classical
generative network from generating the true data efficiently. Furthermore, investigating and generalizing other known variants of generative adversarial networks to the quantum mechanical regime would also be fascinating. This would include such adversarial networks as convolutional, conditional, bidirectional, and semi-supervised.

In conclusion, we have introduced a quantum mechanical generalization of a generative adversarial network, known as a QuGAN. Such a quantum adversarial learning game consists of a generator and a discriminator where the generator is working to trick the discriminator into passing off fake data as real data. In the case of the quantum game converging, the generator generates the same statistics as the true data. Due to the inherent probabilistic nature of quantum mechanics, the proof of the quantum case is simpler than the classical case. 

Finally, we introduced three versions of QuGANs in this work based on whether the real data, fake data, discriminator and generator are quantum mechanical or classical. In the case where the real data is purely classical and high dimensional, and the generator and discriminator are both quantum, we find that quantum adversarial networks potentially exhibit an exponential
advantage over classical adversarial networks. 

\acknowledgements

We thank Pierre-Luc Dallaire-Demers and Nathan Killoran for helpful discussions and suggestions.

\end{document}